# A Unified Shear-Thinning Treatment of Both Film Thickness and Traction in EHD

Scott Bair
Georgia Institute of Technology
Center for High Pressure Rheology
George W. Woodruff School of Mechanical Engineering
Atlanta, GA 30332-0405, USA

and

Philippe Vergne and Michel Querry
Laboratoire de Mécanique des Contacts et des Solides
UMR CNRS/INSA 5514, INSA de Lyon bât. Jean d'Alembert
69621 Villeurbanne cedex, France

May, 2004

**Abstract**

A conclusive demonstration has been provided that the nature of the shear-thinning, that affects both film thickness and traction in EHL contacts, follows the ordinary power-law rule that has been described by many empirical models of which Carreau is but one example. This was accomplished by accurate measurements in viscometers of the shear response of a PAO that possesses a very low critical stress for shear-thinning and accurate measurements in-contact of film thickness and traction under conditions which accentuate the shear-thinning effect. The in-contact central film thickness and traction were entirely predictable from the rheological properties obtained from viscometers using simple calculations.

These data should be invaluable to researchers endeavoring to accurately simulate Hertz zone behavior since the shear-thinning rheology is extensively characterized and accurate in-contact data are available to test. In addition, a new model has been introduced that may be useful for the rheological characterization of mixtures.

1. **Introduction**

The search for an accurate statement for the constitutive behavior of a liquid lubricant under conditions of high pressure and high shear stress has occupied a great portion of the research effort in the field of elastohydrodynamics for more than thirty years. However, an accurate working description of the steady shear rate (or stress) dependence of viscosity, known as shear-thinning, was available early during this period from the pioneering works of such people as Dyson [1] and Hutton [2, 3] and Winer [4]. The generalized viscosity, $\eta$, was known to approach asymptotically the low shear viscosity, $\mu$, as shear rate, $\dot{\gamma}$, (or stress, $\tau$) was decreased through a terminal regime. For high shear rate (or stress) the relationship between viscosity and shear rate and shear stress followed a power-law [1, 2].

$$\frac{\partial \log \tau}{\partial \log \dot{\gamma}} = n \tag{1a}$$

$$\frac{\partial \log \eta}{\partial \log \dot{\gamma}} = n - 1 \tag{1b}$$

$$\frac{\partial \log \eta}{\partial \log \tau} = 1 - \frac{1}{n} \tag{1c}$$

This power law regime was sometimes followed by a second Newtonian with viscosity, $\mu_2$, approached asymptotically from above with increasing rate (or stress) [4]. This was a description obtained from rheological measurements performed in viscometers of various types.

A second description of shear-thinning evolved within the EHL community at about the same time. Bell and coworkers [5] adopted the single flow unit form of the Hahn-Eyring model for thixotropy [6] for their description of shear-thinning by ignoring the stress induced transformation to Newtonian behavior. This results in logarithmic dependence of $\tau$ on $\dot{\gamma}$ at high rates. Eyring specifically rejected this adaptation of his model [7], but the greater portion of the EHL community followed the example of Bell. The most influential work is that of Johnson, [8] for example.

The purpose of this paper is to conclusively demonstrate that the nature of the shear-thinning within a lubricating film in a concentrated contact follows the ordinary power-law form that has been described by empirical models [9] that bear the names Carreau, Carreau-Yasuda, Ellis, Cross, Spriggs, Ferry, Rabinowitsch, and Ree-Eyring [9]. This will be accomplished by calculating both film thickness and traction from the measured nonlinear shear response of a liquid with a low critical shear stress for shear-thinning. In addition, a new model is offered that has similar properties but may be more useful for the liquid mixtures that are blended to form practical lubricants. It is hoped that this demonstration will convince the EHL community that continued modeling of shear-thinning as logarithmic will not advance the field. This goal is made all the more important by the recent development of accurate non-equilibrium molecular dynamic simulations [10, 11] that can investigate the shear response of liquids which have not yet been synthesized under conditions that can not yet be attained with viscometers.

The conditions for the present experimental measurements were selected to enhance the effects of shear-thinning. Rheological measurements were performed at the Center for High Pressure Rheology at Georgia Tech and contact measurements of film thickness and traction were performed at the Laboratoire de Mécanique des Contacts et des Solides at INSA de Lyon, thus combining what are believed to be the best available capabilities for measurements in and out of contact.

## 2. Rheological Measurements

The experimental liquids employed in this study are all polyalphaolefin, PAO. First, the PAO-650 was one of a series of very high viscosity PAOs received from Mobil Corp. This liquid is the subject of the present film thickness and traction study as well. The PAO-100 was studied previously for the effect of shear-thinning on film thickness [12]. The PAO-40 is a product of Exxon-Mobil known as SHF403.

The low shear viscosities of these liquids were measured in falling body viscometers using applied shear stress of $\tau$ = 33 to 65 Pa and are listed in Table 1, along with the pressure-viscosity coefficient defined as

$$\alpha = \left[ \int_o^\infty \frac{\mu(0)}{\mu(p)} dp \right]^{-1}. \tag{2}$$

Sufficiently high pressure was generated to obtain an inflection in the log viscosity versus pressure behavior in the PAO-650 and PAO-40, a feature that is necessary for the unique determination of free-volume parameters.

Two pressurized Couette viscometers were utilized to obtain the shear dependent relative viscosity, $\eta/\mu$, of PAO-650 shown in Figure 1. The 350 MPa viscometer has been described previously [13]. The 700 MPa viscometer is new and is similar to the lower pressure instrument with the following exceptions: The thrust bearing that supports the rotating shaft has been moved to outside of the high pressure environment to increase the pressure capability. The pressurizing fluid and the torque transducer damping fluid are now the same, a perfluorinated hydrocarbon known as FC77 that is immiscible in organic liquids and remains very fluid at high pressure. The pressurizing fluid has a mass density much greater than that of the sample and is admitted to the bottom of the viscometer to compress the sample at the top of the high pressure chamber, thereby eliminating the need for a separate volume make-up cylinder. The sample volume is 5 ml.

The flow curves in Figure 1 show good agreement between the two viscometers. Four combinations of temperature and pressure were utilized as shown in the figure. For the elevated pressure data, either of two models can accurately describe the results. The Carreau [14] equation is preferred for our traction calculations where shear rate is the independent variable, since in that case shear rate is assumed constant within the Hertz region.

$$\eta = \mu \left[ 1 + (\lambda \dot{\gamma})^2 \right]^{(n-1)/2} \tag{3}$$

A modification [15] of the Carreau equation is preferred for film thickness calculations where shear stress is the independent variable, since the form of the shear stress distribution across the film is known.

$$\eta = \left[1 + \left(\frac{\tau}{G}\right)^2\right]^{(1-1/n)/2} \qquad (4)$$

The two models can be related by writing the characteristic time as a relaxation time

$$\lambda = \mu/G \qquad (5)$$

where $G$ is an effective shear modulus. For power law exponent, $n > 0.2$ equations (3) and (4) yield roughly the same result [9]. For the elevated pressures in Figure 1, $n = 0.74$ and $G = 3.1 \times 10^4$ Pa reasonably describe the shear-thinning shown. Making $G$ a function of the low shear viscosity, $\mu$, can improve the fit [12] over large intervals of viscosity, but for the contact measurements in the present work, the viscosities in contact are similar to those in the Couette viscometers.

An interesting effect is shown in Figure 1 for atmospheric (0.1 MPa) pressure. The measured viscosity diverges from the trend obtained at elevated pressures for shear stress greater than about $3 \times 10^4$ Pa. This is believed to be one of the forms of liquid failure, known as shear cavitation, that has previously been reported and studied [16]. This failure occurs when a principal normal stress becomes tensile. Flowing liquids do not easily support tension [17] and cavitate.

The upper-convected Maxwell model gives for the first normal stress difference [18]

$$N_1 = 2G\lambda^2 \dot{\gamma}^2 \qquad (6)$$

This value represents an extra tensile stress in the direction of flow as compared to the normal stress in the cross-flow direction. Then the maximum principal tensile stress is from elementary stress analysis

$$\sigma_1 = -p + \frac{1}{2}\left(N_1 + \sqrt{N_1^2 + 4\tau^2}\right) \qquad (7)$$

Shear cavitation should occur for $\sigma_1 > 0$, that is to say, when a tensile stress appears in the liquid. For the PAO-650 at $p = 0.1$ MPa, atmospheric pressure, using equations (6) and (7), the tensile stress appears at $\tau > 3.3 \times 10^4 \, Pa$ and this represents a limit for the measurement of constitutive behavior as can be seen in Figure 1. This cavitation mechanism may be important to lubrication and has been studied at various laboratories [16].

The viscosity of PAO-650 is plotted versus shear stress in Figure 2 for $p = 100$ MPa and $T = 20°C$. The curve shown in Figure 2 is the modified Carreau equation (4). The usual Carreau equation (3) is indistinguishable from equation (4) on this graph. Also shown in Figure 2 are two flow curves for PAO-100 and two flow curves for a blend of 20% weight PAO-650 in PAO-100. Later in this paper, a new model will be introduced to describe mixtures, and this model was used to plot the curves in Figure 2. Notice that the flow curve for the mixture displays two shear thinning transitions separated by a "plateau" that might be roughly described as a second Newtonian. In the next two sections, film thickness and traction will be calculated and compared to measurements on PAO-650 using equations (3) and (4).

### 3. Traction and Film Thickness Measurements

The concentrated contact lubrication measurements reported here were performed with a steel ball of 12.7 mm radius loaded at 32N against the face of a glass disc. The combined elastic modulus is 123.9 GPa, yielding a Hertz pressure of $p_H = 528$ MPa. The specimens are driven by independent motors to produce the desired slide to roll ratio. The stability of the ball and disk velocities is controlled with high precision.

The ball and the disk used in this study were carefully polished leading to a composite RMS roughness of the undeformed surfaces lower than 5 nm. The glass disk was coated on its underside with a thin semi-reflective chromium layer that was overlaid by a silicon dioxide spacer layer of the same refractive index as the studied lubricant. This technique has been pioneered by Westlake et al. [19] at the end of the nineteen sixties to overcome the major limitations to the classical optical interferometry technique. In this work the spacer layer was 65 nm thick.

The bottom of the ball dips into a reservoir containing the lubricant, ensuring fully flooded conditions in the conjunction. The contact and the lubricant are thermally isolated from the outside and heated (or cooled) by an external thermal controlling system. A platinum temperature probe monitors the lubricant temperature in the test reservoir. The temperature for all tests was 75°C.

The film thickness measurement technique used in this work is based on differential colorimetric interferometry and has been detailed in refs. [20, 21]. The contact area is illuminated with a halogen white light source. The chromatic interferograms produced by the contact are captured by the 3CCD color video camera and frame-grabbed by a personal computer. The spatial resolution of the captured images is high, about 1.1 $\mu m$ compared to the contact diameter of 340 $\mu m$.

Simultaneously during film thickness measurements, traction forces and normal load are recorded by a new multi-axis strain gauge sensor. It combines a broad range of measurable forces, appropriate sensitivities over the different directions, and a high stiffness.

Measurements of traction coefficient as a function of slide-to-roll ratio, $\Sigma$, are plotted as the data points in Figure 3 for three different values of rolling velocity. The values of traction coefficient for Figure 3 are unusually small (< 0.015) and near the threshold of resolution of the instrument at low slide-to-roll ratio. These low traction values apparently result from the low Newtonian limit of the liquid.

Measurements of central and minimum film thickness are plotted as the data points in Figures 4 and 5, respectively for pure rolling, $\Sigma = 0$. The central film thickness, in Figure 4, was measured to be 72% of the Newtonian prediction of Hamrock and Dowson [22] at low speed decreasing to 52% at high speed. The minimum film thickness, shown in Figure 5, remained at a nearly constant 55% of the Hamrock and Dowson [22] Newtonian prediction. This minimum film behavior is unexpected and cannot be investigated through numerical simulation at present. Measurements of the effect of sliding on film thickness for two rolling velocities are shown as the data points in Figure 6. Central film thickness is insensitive to sliding, within the operating conditions imposed during these experiments.

## 4. Traction Calculations

The calculation of viscous traction employed here is very simple. The contribution from roller and disc elastic compliance can be neglected since the contact pressure is low. The traction gradient due to elastic compliance is of the order of $G_s/p_H$ [23] where $G_s$ is the shear modulus of the solids. If the combined effective shear modulus of the solids is as low as $G_s = 10$ GPa, then the traction gradient due to roller/disc compliance is of the order of 20 which is much greater than the largest gradient in Figure 3 for Newtonian response, about 0.2. Then the contribution from the roller and disc is much stiffer than the liquid film.

The variation of low shear viscosity for PAO-650 with pressure for all calculations is represented by the Doolittle [24] equation,

$$\mu(p) = \mu_o \exp\left[ B \frac{v_{occ}}{v_o} \left( \frac{1}{\left(\frac{v}{v_o} - \frac{v_{occ}}{v_o}\right)} - \frac{1}{\left(1 - \frac{v_{occ}}{v_o}\right)} \right) \right] \qquad (8)$$

where $v$ is volume, $v_o$ is volume at $p = 0$, and $v_{occ}$ is the occupied volume that is independent of pressure. The variation of volume with pressure is described with the Tait equation [24].

$$v/v_o = 1 - \frac{1}{K_o' + 1} \ln\left[1 + \frac{p}{K_o}(1 + K_o')\right] \qquad (9)$$

A least squares regression of viscosity data for 75°C resulted in $K_o = 1.425$ GPa, $K_o' = 12.82$, $v_{occ}/v_o = 0.6694$, $B = 4.422$ and $\mu_o = 1.42$ Pa·s.

The shear rate in the Hertz region can be calculated from

$$\dot{\gamma} = \frac{u\Sigma}{h_c} \qquad (10)$$

assuming a uniform value of $h_c$ over the Hertz contact area.

The central film thickness, $h_c$, was obtained by the method of the next section setting $h_c = h_{NN}$. The average contact shear stress results from integration of the local stress, $\tau$, over the contact area,

$$\bar{\tau} = \int_o^1 2r\tau dr \tag{11}$$

for dimensionless contact radius, $0 < r < 1$. For point contact the average pressure is $\bar{p} = 2/3\, p_H$ and the traction coefficient is simply $\bar{\tau}/\bar{p}$. We assume that the pressure distribution is identical to that of the unlubricated Hertz contact.

$$p = p_H \left(1 - r^2\right)^{\frac{1}{2}} \tag{12}$$

Traction is calculated for the Newtonian and Carreau cases where

$$\tau = \mu(p)\dot{\gamma} \tag{13}$$

and

$$\tau = \eta(p, \dot{\gamma})\dot{\gamma} \tag{14}$$

respectively. Equation (3) is used for $\eta$ in equation (14). The traction curves for these cases are indicated in Figure 3.

The Newtonian case, as expected, greatly overestimates the traction coefficient, $\bar{\tau}/\bar{p}$. The shear-thinning calculation is remarkably accurate as indicated by the Carreau curves in Figure 3. For these results, $G$ is assumed constant. Limiting the shear stress to 0.035p had no effect on the traction calculations. As in a previous calculation for a low pressure contact [23] the local shear stress may not reach the limiting value.

## 5 Film Thickness Calculations

The Hamrock and Dowson [22] Newtonian central film thickness, $h_N$, may be corrected for shear-thinning by the simple formula by Bair and Winer [25] for the non-Newtonian central film thickness, $h_{NN}$.

$$\phi = h_N / h_{NN} = \left[1 + 4.44\left(\frac{u\mu_o}{h_N G}\right)^{1.69}\right]^{1.26(1-n)^{1.78}} \quad (15)$$

where $u$ is the absolute value of the rolling velocity. In practice, the film thickness is found by dividing any Newtonian based calculation of central thickness, $h_N$, by $\phi$, a number greater than or equal to one. Equation (15) was obtained by computing the reduction of film thickness due to shear-thinning in a Grubin style line contact numerical solution with the modified Carreau equation (4), assuming $G$ constant. For large rolling velocity, $u$, and using equation (15) to correct the Hamrock and Dowson equation, film thickness will vary with rolling velocity as

$$h_{NN} \propto u^{[0.67 - 0.703(1-n)^{1.78}]} \quad (16)$$

For PAO-650, $n = 0.74$ and film thickness should vary with speed to the 0.606 power which is a good approximation to the gradient of the data in Figure 4 as opposed to the Newtonian gradient of 0.67.

Applying the correction formula, equation (15), to the Newtonian prediction in Figure 4 results in the corrected curve in the figure. Clearly, correcting for the measured shear dependence of viscosity results in improved central film thickness calculations. For the measured minimum film thickness shown in Figure 5, however, the slope is about the same as for the Newtonian prediction. Then the minimum thickness must become more sensitive to shear-thinning than the central thickness when the film is thin. An explanation of the minimum film thickness behavior in Figure 5 would require a "full" numerical contact simulation including the Hertz region. Such calculations cannot presently be performed with realistic models, but it seems reasonable that enhanced side leakage due to shear-thinning could cause additional film thinning at the minimum film locations.

Calculations of the effect of sliding on film thickness were performed with the numerical scheme described in reference [26]. This is again a Grubin style line contact inlet zone analysis. The pressure variation of low shear viscosity was represented by equations (8) and (9). The shear dependent viscosity was represented by equation (4). Results are shown as the curves in Figure 6 along with the Hamrock and Dowson Newtonian prediction. Clearly, the shear-thinning line contact calculation is more accurate for this point contact application than is the Newtonian point contact formula. The calculated film thickness is slightly more sensitive to sliding than is the measured film thickness although the difference in sensitivity is small. In previous work [27] the authors found the opposite trend; the prediction was less sensitive to sliding than the measurement. Sliding increases the magnitude of shear along one solid boundary while decreasing the magnitude of shear along the other boundary for low $\Sigma$ and so for low $\Sigma$ the sliding effect is not great for ordinary shear-thinning.

## 6. A New Shear-Thinning Model

The demonstrations of the effects of shear-thinning on traction and film thickness given above emphasize the need for accurate modeling of the rheology in practical concentrated contact problems. Real lubricants are generally mixtures of species of varying molecular weight. A common example is that of a multigrade motor oil with a polymeric VI improver. These commonly display flow curves with a second plateau [4] such as the mixture of PAO-650 + PAO-100 in Figure 2. The plateau must end where shear-thinning of the solvent base oil begins as in Figure 2.

A second Newtonian response with viscosity, $\mu_2$, is generally accommodated in a shear-thinning model by multiplying the model by the difference in the two Newtonians so that equation (4), for example, would read

$$\eta = \mu_2 + (\mu - \mu_2)\left[1+\left(\frac{\tau}{G}\right)^2\right]^{(1-1/n)/2}. \qquad (17)$$

Let

$$F_i(\tau) = \left[1 + \left(\frac{\tau}{G_i}\right)^2\right]^{(1-1/n_i)/2} \quad (18)$$

and $x_1 = 1 - \mu_2/\mu$ and $x_2 = \mu_2/\mu$. Then equation (17) may be obtained by writing

$$\eta = \mu \sum_{i=1}^{N} x_i F_i(\tau) \quad (19)$$

if $G_2 \to \infty$ and $N = 2$. Equation (19) should be immediately recognized as the Ree-Eyring model for shear-thinning [7], that invokes the concept of multiple flow units in order to construct a flow curve, with an important difference in $F_i(\tau)$. Eyring used a series of inverse hyperbolic sine functions of shear rate so that the high shear behavior resulted in a logarithmic variation of stress with rate. Since this is not accurate at high shear rates [7], Eyring required multiple flow units, $N > 1$, for even monodisperse liquids. The proposed new model, equation (19), overcomes this difficulty by employing a function that naturally follows a power-law at high shear. The full model is

$$\eta = \mu \sum_{i=1}^{N} x_i \left[1 + \left(\frac{\tau}{G_i}\right)^2\right]^{(1-1/n_i)/2}. \quad (20)$$

The parameters of the new model are listed in Table 2 for the liquids shown in Figure 2 and for the PAO-40. The curves shown in Figure 2 represent the new model. For the PAO-650 and PAO-40, $N = 1$ and equation (20) is identical to equation (4). The other "straight cut" base oil, PAO-100, required two flow units because of a broad transition as seen in Figure 2 that is often accommodated by the Yasuda modification to the Carreau equation [9]. This broad transition is apparently due to the presence of a very high molecular weight fraction that was detected in the z-moment molecular weight by gel permeation chromatography.

An advantage obtained from the use of this new model is evident for the mixture of PAO-650 and PAO-100 when the values of $G_i$ are examined in Table 2. These may be thought of as critical stresses for shear-thinning transitions and each $G_i$ of the mixture is found in the components. One comes from PAO-650 and two come from PAO-100. The power law exponents, $n_i$, are however different. It is possible that mixing rules for shear-thinning might be constructed, from much more data of course, that would be important for lubricant blending. For dilute solutions of polymer in low molecular weight base oils the shear stress of the first shear-thinning transition is known from both theory and experiment [12] to vary in proportion to polymer concentration so that there must be a limit to the usefulness of the above generalization regarding $G_i$. Also, in the limit of high shear stress (or rate) the term with the largest value of $n_i$ dominates so that the effective power-law index at very high shear will become the largest $n_i$. It is, of course, possible that this will be correct but there is no data at this time to support this type of high shear behavior.

## 7. **Conclusions**

A conclusive demonstration has been provided that the nature of the shear-thinning, that affects both film thickness and traction in EHL contacts, follows the ordinary power-law rule that has been described by many empirical models of which Carreau is but one example. This was accomplished by accurate measurements in viscometers of the shear response of a PAO that possesses a very low critical stress for shear-thinning and accurate measurements in-contact of film thickness and traction under conditions which accentuate the shear-thinning effect. The in-contact central film thickness and traction were entirely predictable from the rheological properties obtained from viscometers using simple calculations.

These data should be invaluable to researchers endeavoring to accurately simulate Hertz zone behavior since the shear thinning rheology is extensively characterized and accurate in-contact data are available to test. In addition, the Hertz pressure for the in-contact measurements is sufficiently low and the traction is sufficiently low that the product of the local shear stress and the local pressure viscosity coefficient is always much less than one. This is a criterion for validity of the Reynolds equation which is generally not valid for EHL [28, 29].

In addition, a new model has been introduced that may be useful for the rheological characterization of mixtures.

**Acknowledgements**

One of the authors (Scott Bair) is thankful for the support of the Timken Company. Margaret Wu of ExxonMobil provided the GPC analysis of PAO-100.

### Nomenclature

| | |
|---|---|
| $\mu$ | low shear viscosity, Pa·s |
| $\eta$ | generalized viscosity, Pa·s |
| $\tau$ | shear stress, Pa |
| $\dot{\gamma}$ | shear rate, s$^{-1}$ |
| $n$ | power-law exponent |
| $G$ | modulus or critical stress, Pa |
| $u$ | absolute value of rolling velocity, m/s |
| $\mu_o$ | low shear viscosity at ambient pressure, Pa·s |
| $h$ | measured film thickness, m |
| $h_N$ | film thickness by a Newtonian calculation, m |
| $h_{NN}$ | film thickness corrected for shear-thinning, m |
| $\phi$ | $h_N/h_{NN}$ |
| $N$ | number of flow units |
| $x_i$ | weighting factor |
| $R$ | reduced disc radius, m |
| $T$ | temperature, °C |
| $a$ | Yasuda parameter |
| $\lambda$ | characteristic time, s |
| $\Sigma$ | slide-to-roll ratio |
| $\mu_2$ | viscosity of a second Newtonian, Pa·s |
| $\alpha$ | pressure-viscosity coefficient, Pa$^{-1}$ |
| $N_1$ | first normal stress difference, Pa |
| $\sigma_1$ | greatest principle tensile stress, Pa |
| $v$ | volume, m$^3$ |
| $v_o$ | volume at p = 0, m$^3$ |
| $v_{occ}$ | occupied volume, m$^3$ |
| $p$ | pressure, Pa |
| $K_o$ | bulk modulus at $p = 0$, Pa |

$K_o'$          pressure derivative of bulk modulus at $p = 0$

r          dimensionless contact radius

Table 1. Low Shear Viscosity, $\mu$, in Pa·s

| p/MPa | PAO-650 20°C | PAO-650 50°C | PAO-650 75°C | PAO-100 20°C | 20% PAO-650 in PAO-100 20°C | PAO-40 25°C |
|---|---|---|---|---|---|---|
| 0.1 | 21.9 | 3.94 | 1.42 | 4.23 | 5.66 | 0.752 |
| 50 | 64.3 | 9.97 | 3.30 | 13.0 | 17.8 | 2.17 |
| 100 | 164 | 22.2 | 6.78 | 35.5 | 49.2 | 5.68 |
| 150 | --- | --- | --- | 88.2 | 119 | --- |
| 200 | 889 | 89.5 | 22.4 | 199 | 262 | 30.1 |
| 250 | --- | --- | --- | 427 | 562 | --- |
| 300 | 4060 | 284 | 61.8 | 890 | 1184 | 134 |
| 350 | --- |  | --- | 1840 | 2452 | --- |
| 400 | 17420 |  | --- | 3860 |  | 506 |
| 450 |  |  | 292 |  |  | --- |
| 500 |  |  | --- |  |  | 1726 |
| 600 |  |  | 1020 |  |  | 6350 |
| 700 |  |  | --- |  |  | 23000 |
| 750 |  |  | 2770 |  |  | --- |
| 800 |  |  | --- |  |  | 90400 |
| 900 |  |  | 9730 |  |  |  |
| 1000 |  |  | 22000 |  |  |  |
| $\alpha/GPa^{-1}$: | 20.1 | 16.8 | 14.8 | 21.2 | 21.5 | 20.0 |

## Table 2. Parameters for the New Rheological Model

$$\eta = \mu \sum_{i=1}^{N} x_i \left[ 1 + \left( \frac{\tau}{G_i} \right)^2 \right]^{(1-1/n_i)/2}$$

| Fluid | N | i | $x_i$ | $n_i$ | $G_i$/Pa |
|---|---|---|---|---|---|
| PAO-650 | 1 | 1 | 1.0 | 0.74 | 3.1 x $10^4$ |
| PAO-100 | 2 | 1 | 0.5 | 0.8 | 1 x $10^5$ |
|  |  | 2 | 0.5 | 0.5 | 4 x $10^6$ |
| 20% PAO-650 in PAO-100 | 3 | 1 | 0.4 | 0.28 | 3.1 x $10^4$ |
|  |  | 2 | 0.3 | 0.75 | 1 x $10^5$ |
|  |  | 3 | 0.3 | 0.4 | 4 x $10^6$ |
| PAO-40 | 1 | 1 | 1.0 | 0.54 | 3.8 x $10^6$ |

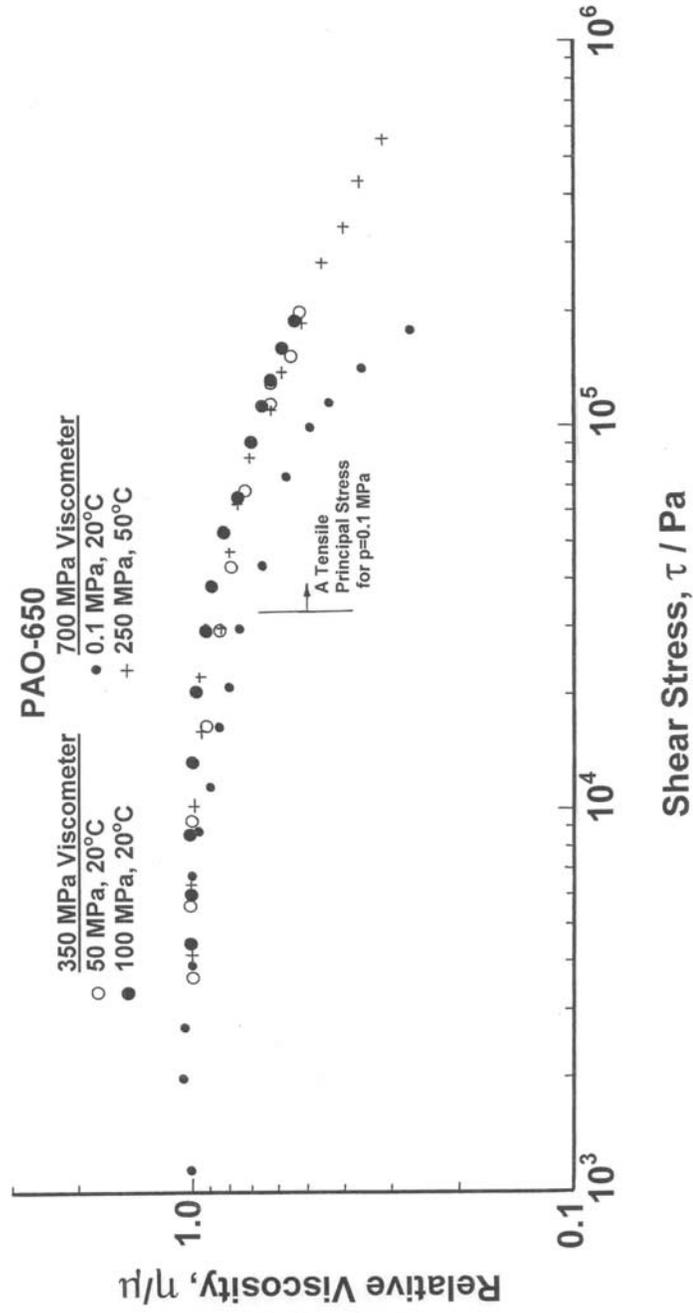

Figure 1. The shear dependent viscosity of PAO-650 using two pressurized Couette viscometers with shear cavitation occurring at atmospheric pressure.

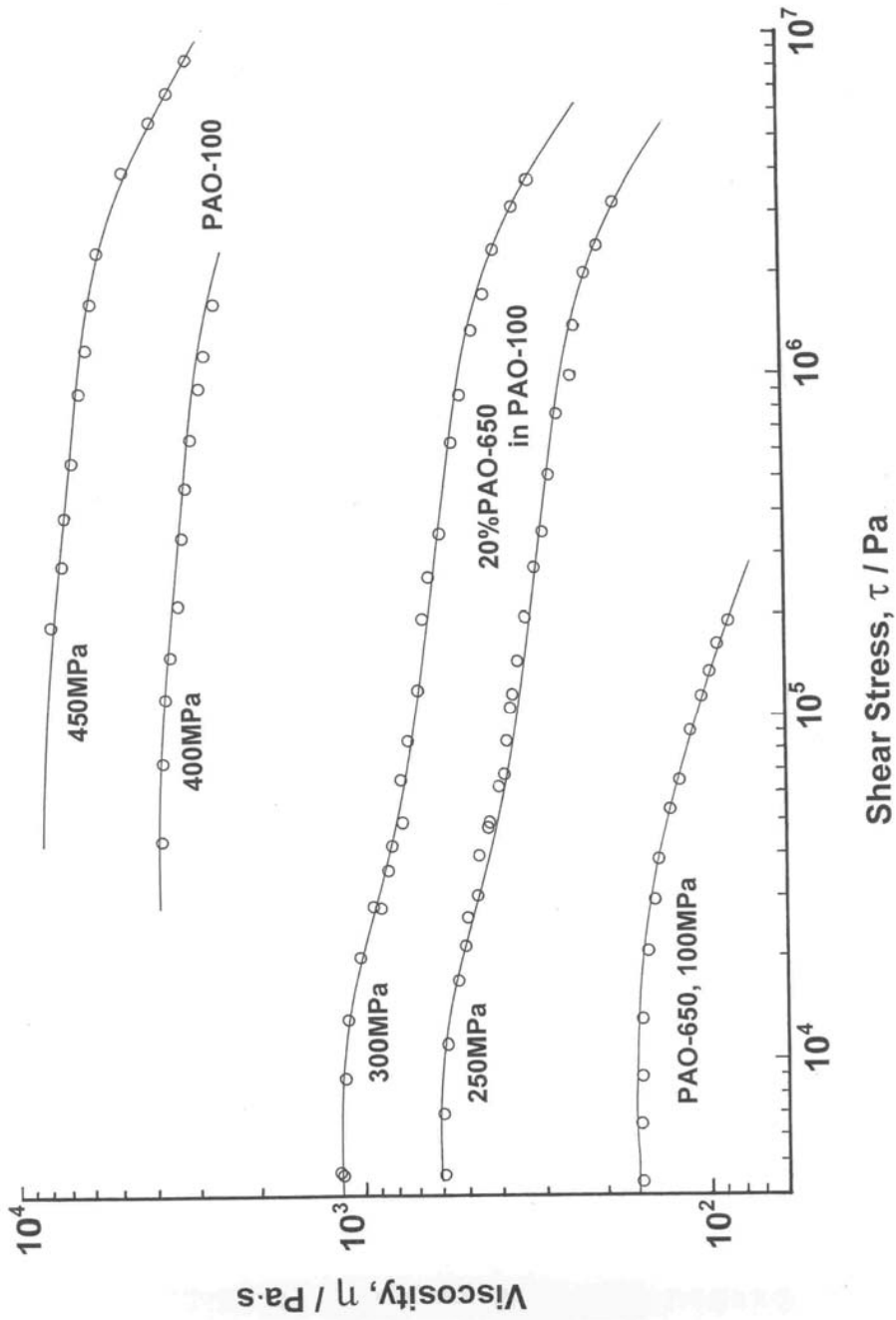

Figure 2. The shear dependent viscosities of PAO-650, PAO-100 and a mixture of PAO-650 + PAO-100 for T = 20°C

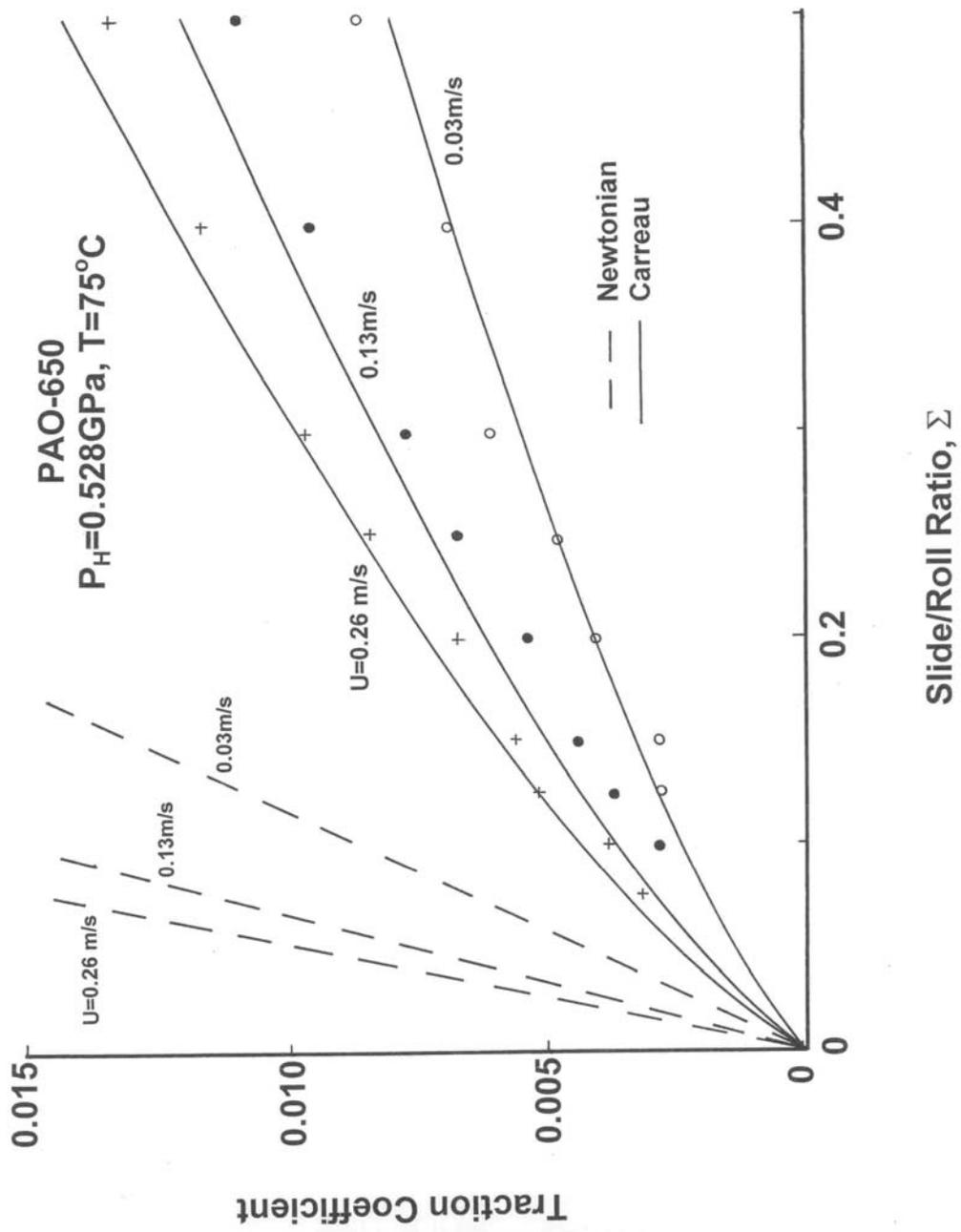

Figure 3. Traction curves for PAO-650. Points are measurements.

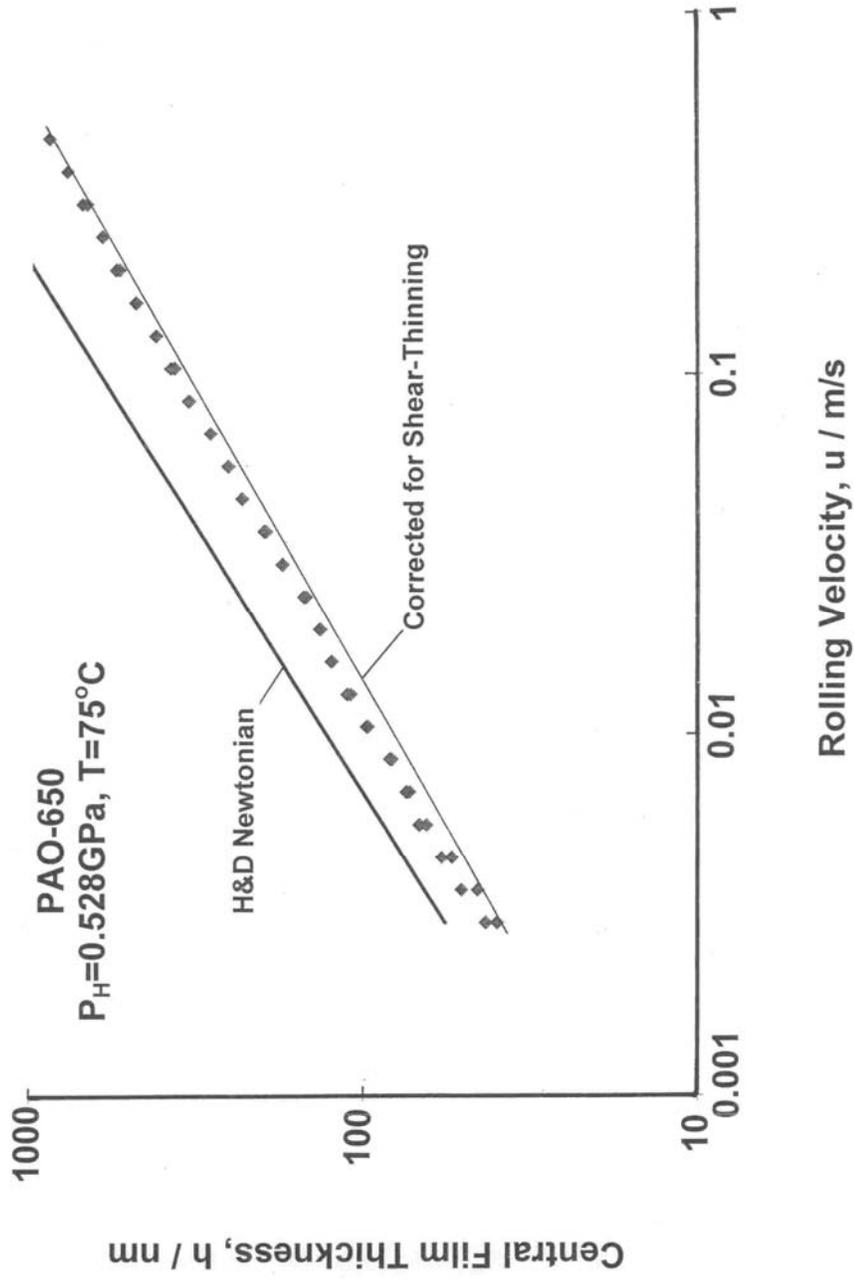

Figure 4. Central film thickness for PAO-650 in pure rolling. Points are measurements.

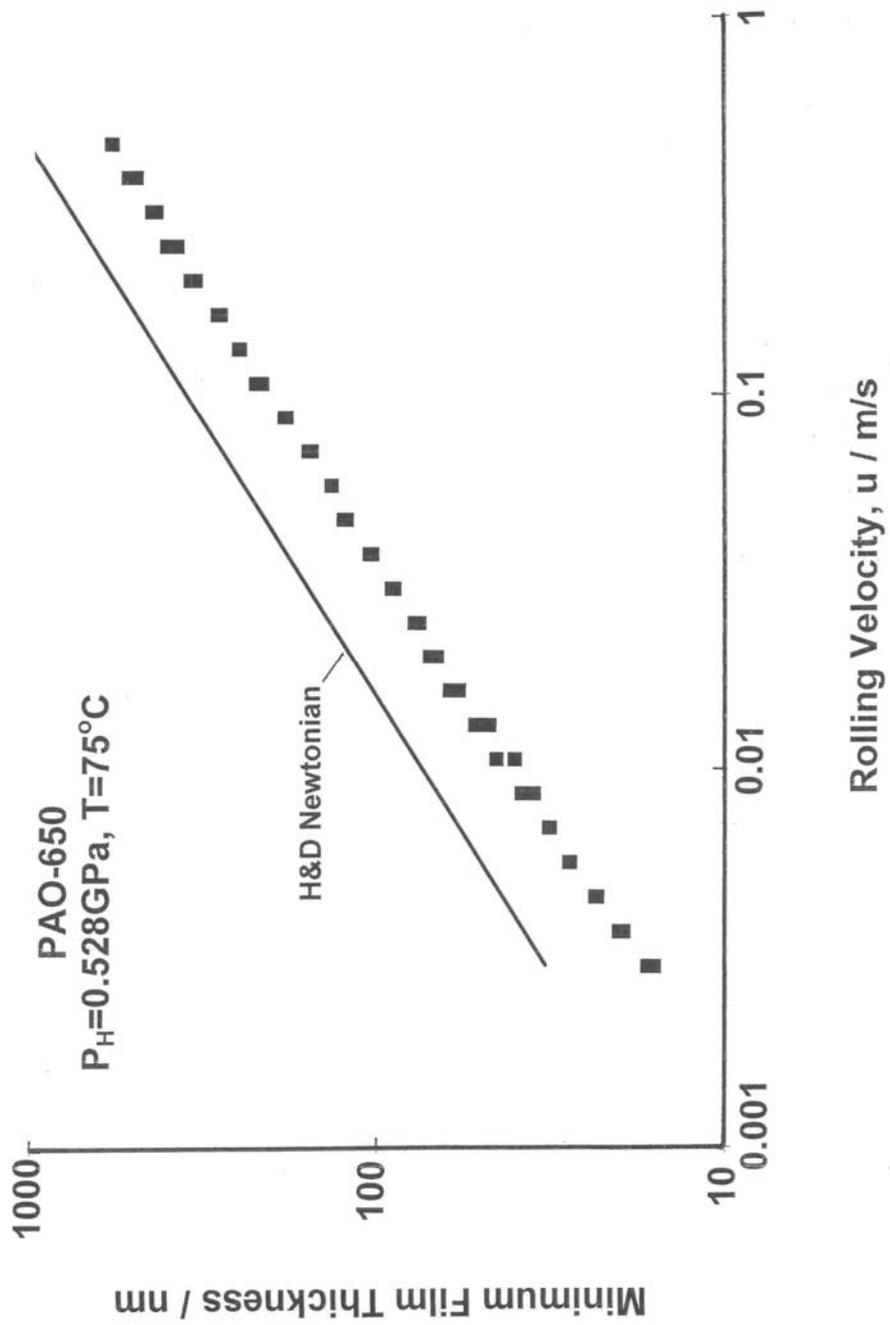

Figure 5. Minimum film thickness for PAO-650 in pure rolling. Points are measurements.

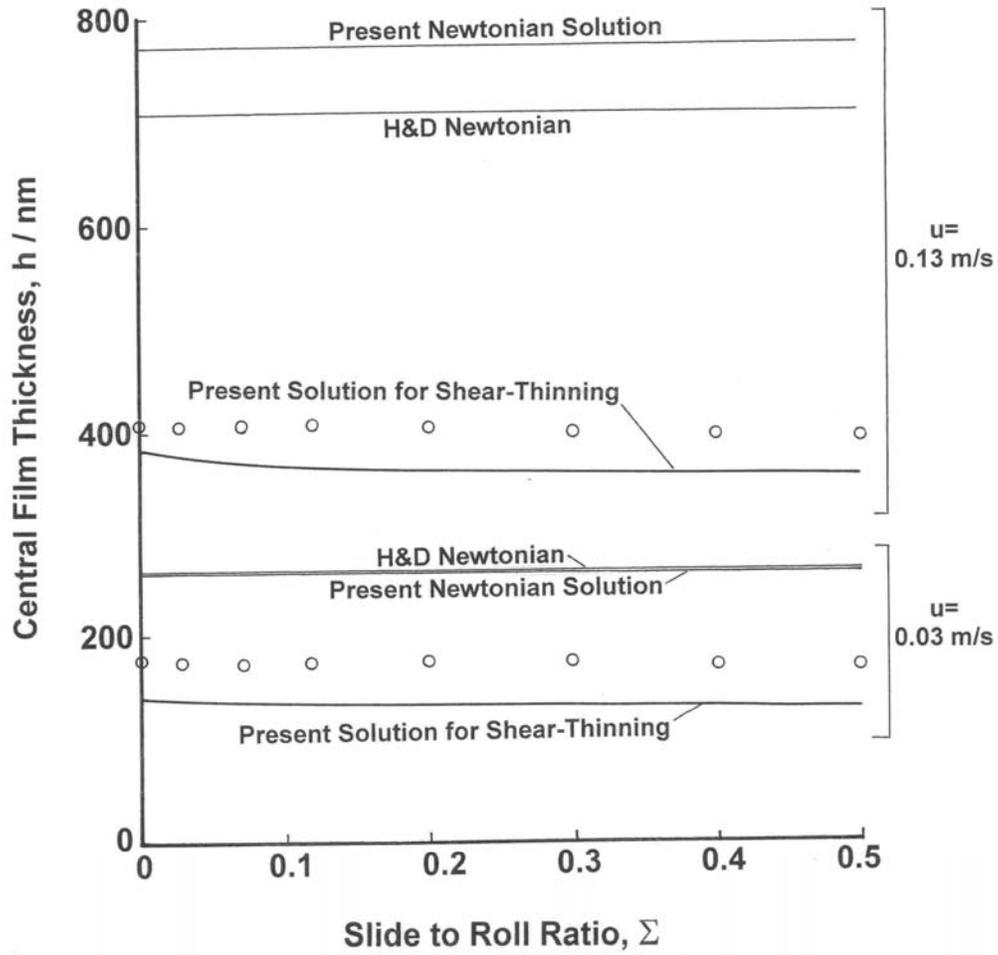

**Figure 6. Central film thickness for sliding contact. Points are measurements.**